\def\sR{\mathcal{R}}

\documentclass[12pt,preprint]{aastex}
\shorttitle{Shock Transition Zone}
\shortauthors{Heng \& McCray}
\begin{document}

\title{THE TRANSITION ZONE IN BALMER-DOMINATED SHOCKS}

\author{Kevin Heng\altaffilmark{1,2,3}, Matthew van Adelsberg\altaffilmark{1}, Richard McCray\altaffilmark{1}, John C. Raymond\altaffilmark{4}}

\altaffiltext{1}{JILA, University of Colorado, 440 UCB, Boulder, CO 80309; hengk@colorado.edu}

\altaffiltext{2}{Max Planck Institut f\"{u}r Extraterrestrische Physik, Giessenbachstra$\beta$e, 85478 Garching, Germany}

\altaffiltext{3}{Max-Planck-Institut f\"{u}r Astrophysik, Karl-Schwarzschild-Strasse 1, 85740 Garching, Germany}

\altaffiltext{4}{Harvard-Smithsonian Center for Astrophysics, 60 Garden Street, Cambridge, MA 02138}

\begin{abstract}
We examine the structure of the post-shock region in supernova remnants (SNRs).  The ``shock transition zone'' is set up by charge transfer and ionization events between atoms and ions, and has a width $\sim 10^{15}$ cm$^{-2}$ $n^{-1}_0$, where $n_0$ is the total pre-shock density (including both atoms and ions).  For Balmer-dominated SNRs with shock velocity $v_s \gtrsim 1000$ km s$^{-1}$, the Rankine-Hugoniot conditions for ion velocity and temperature are obeyed instantly, leaving the full width at half-maximum (FWHM) of the broad H$\alpha$ line versus $v_s$ relation intact.  However, the spatial variation in the post-shock densities is relevant to the problem of Ly$\alpha$ resonant scattering in young, core-collapse SNRs.  Both two- (pre-shock atoms and ions) and three-component (pre-shock atoms, broad neutrals and ions) models are considered.  We compute the spatial emissivities of the broad ($\xi_b$) and narrow ($\xi_n$) H$\alpha$ lines; a calculation of these emissivities in SN 1006 is in general agreement with the computed ones of Raymond et al. (2007).  The (dimensionless) spatial shift, $\Theta_{\rm{shift}}$, between the centroids of $\xi_b$ and $\xi_n$ is unique for a given shock velocity and $f_{\rm{ion}}$, the pre-shock ion fraction.  Measurements of $\Theta_{\rm{shift}}$ can be used to constrain $n_0$.
\end{abstract}


\section{INTRODUCTION}
\label{sect:intro}

In a purely hydrodynamical treatment, shock fronts are regarded as mathematical discontinuities, across which the density, pressure and temperature of the fluid vary according to the Rankine-Hugoniot jump conditions.  A characteristic length scale for the thickness of the discontinuity appears when one takes into account the atomic structure of the gas (Zel'dovich \& Raizer 1966).  The structure of collisional shock fronts is relatively well understood.  A dissipation mechanism is generally required: for example, thermal conduction and viscosity for weak and strong shocks, respectively.  In {\it collisionless} shock fronts, the  dissipation is due to turbulent electromagnetic fields rather than collisions.  Determining the structure of collisionless shocks requires an understanding of how particles equilibrate their temperatures via plasma instabilities, which is currently a largely open question in astrophysics.  Reviews can be found in McKee \& Hollenbach (1980) and Draine \& McKee (1993).

\begin{figure}
\begin{center}
\includegraphics[width=6in]{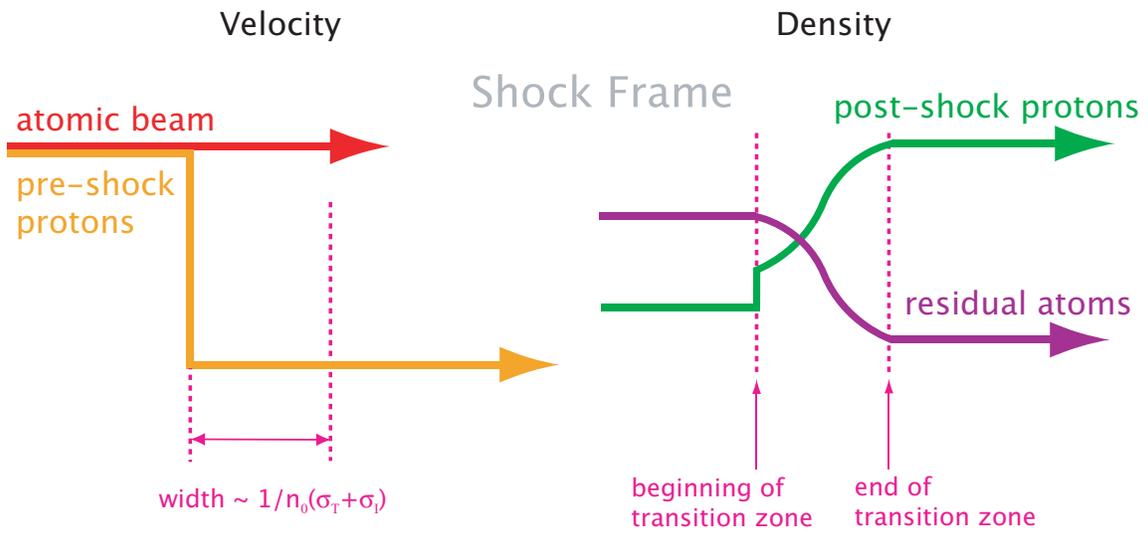}
\end{center}
\caption{Schematic diagram of the shock transition zone, in the case of a strong shock.  The width of the transition zone is on the order of the mean free path of interactions (charge transfer and ionization).  The velocity of the ions goes down to 1/4 of its pre-shock value almost immediately, according to the Rankine-Hugoniot jump condition.  The ion density first jumps by a factor of 4 to conserve momentum, then eventually evolves to a value which depends on the pre-shock ion fraction, $f_{\rm{ion}}$.  }
\label{fig:zone}
\end{figure}

Our study is motivated by a need to understand the structure of the post-shock region in Balmer-dominated supernova remnants (SNRs; Heng \& McCray 2007, hereafter HM07, and references therein), which we term the ``shock transition zone''.  These SNRs typically have shock velocities $v_s \sim 1000$ km s$^{-1}$, ages much less than the radiative cooling times and produce a modest amount of ionizing radiation.  Balmer-dominated SNRs have two-component spectra consisting of a broad and a narrow line (Chevalier, Kirshner \& Raymond (1980, hereafter CKR80; see \S\ref{subsect:bdsnr}).  If the temperatures of the atoms and ions are known, the width of the broad line is uniquely related to the shock velocity, as first shown by CKR80 The upstream, pure hydrogen gas consists of atoms and ions ($T_u \sim 10000$ K) with a typical pre-shock ion fraction of $f_{\rm{ion}} \sim 0.5$.  The atoms are converted into ions via charge transfer (with protons) and impact ionization (with both electrons and protons) in the transition zone (Fig. \ref{fig:zone}), which has a width on the order of the mean free path of interactions.  

If a noticeable variation of the ion velocity occurs within the transition zone, then for a given broad line width, the shock velocity would be {\it under-estimated}.  This is because charge transfer events, which give rise to broad H$\alpha$ emission, are favored at low relative velocities between the atoms and ions, which may occur within the transition zone.  In \S\ref{subsect:bdsnr}, we show this {\it not} to be the case.

The second reason for investigating the structure of Balmer-dominated SNRs is that Ly$\alpha$ resonant scattering occurs in the freely streaming debris of young, core-collapse SNRs that are still in the pre-Sedov-Taylor phase.  Photons produced at the shock fronts undergo a spectral random walk and become increasingly redshifted as the debris is in Hubble-like flow.  Ly$\alpha$ line profiles will therefore be distorted with respect to non-resonant lines such as H$\alpha$ (if the hydrogen atoms are mostly in the ground state).  Photon production occurs within the shock transition zone, the width of which is usually greater than the mean free path of the Ly$\alpha$ photons.  Therefore, understanding the spatial structure of the transition zone in Balmer-dominated SNRs is relevant to modeling the Ly$\alpha$ lines in young remnants like SNR 1987A, where resonant scattering occurs in the hydrogen ejecta from the massive progenitor.

The third motivation for our study is to develop a new method for deriving the spatial emissivity profiles of both the broad and narrow H$\alpha$ lines behind the shock front.  Raymond et al. (2007, hereafter R07) have shown that such profiles can be used to infer $n_0$, the total pre-shock density (including both atoms and ions).

In \S\ref{sect:model}, we present our model and assumptions and state the relevant equations involved.  We describe our solution methods in \S\ref{sect:solutions} and analyze our results in \S\ref{sect:results}.  In \S\ref{sect:discussion}, we discuss the implications and limitations of our results and present opportunities for future work.

\section{MODEL \& ASSUMPTIONS: TWO-COMPONENT MODEL}
\label{sect:model}

For the problem to be tractable, we need to make a few simplifying assumptions.  Firstly, the only significant sink term present for the atoms is for their conversion to ions, via charge transfer and impact ionization.  In the frame of the shock, the atoms comprise a beam with a velocity of $v_{\rm{H}}=v_s$.  The ions are shocked and isotropized in a distance comparable to the cyclotron gyroradius ($l_{\rm{gyro}} \sim 10^{10}$ cm).  We assume that the ion distribution function  becomes approximately Maxwellian in a length scale that is much less than the width of the shock transition zone.  Plasma instabilities are capable of partially equilibrating the temperatures between the electrons and protons ($T_e/T_p \ga 0.1$ for $v_s \lesssim 1000$ km s$^{-1}$; Cargill \& Papadopoulos 1988; Rakowski, Ghavamian \& Hughes 2003; Ghavamian, Laming \& Rakowski 2007). We parametrize the ratio of the temperatures at the shock front by
\begin{equation}
\beta = \frac{T_e}{T_p}.
\end{equation}
Hereafter, we drop the ``$p$'' subscript when referring to variables describing the ion fluid.  The length scale for equilibration is determined by Coulomb collisions and can be much larger than the physical extent of the SNR for $\beta=0.1$.  It is generally larger than $l_{\rm{gyro}}$ and the length scale for atomic interactions, $l_{\rm{zone}} \sim 10^{15}$ cm$^{-2}$ $n^{-1}_0$.

\subsection{INTERACTION OF ATOMIC BEAM WITH MAXWELLIAN IONS}
\label{subsect:equations}

In the shock transition zone, the atomic beam (number density $n_{\rm{H}}$ and velocity $v_{\rm{H}}$) interacts with Maxwellian ions (density $n$, velocity $v$ and temperature $T$).  Charge transfers between the atomic beam and post-shock protons produce populations of atoms having velocity distributions intermediate between a beam and a Maxwellian, which we term ``broad neutrals''\footnote{In this paper, we refer to the pre-shock atoms in a beam simply as ``atoms'', while ``broad neutrals'' refer to the post-shock atoms found in broad distributions.  Alternatively, the ``atoms'' could have been named ``narrow neutrals''.  We have avoided use of the terms ``slow'' and ``fast neutrals'' as these are dependent on one's frame of reference.}.  The resulting ``skewed Maxwellian'' distributions are described in HM07.  These broad neutrals are produced in the same transition zone as the ions.  

\begin{figure}
\begin{center}
\includegraphics[width=6in]{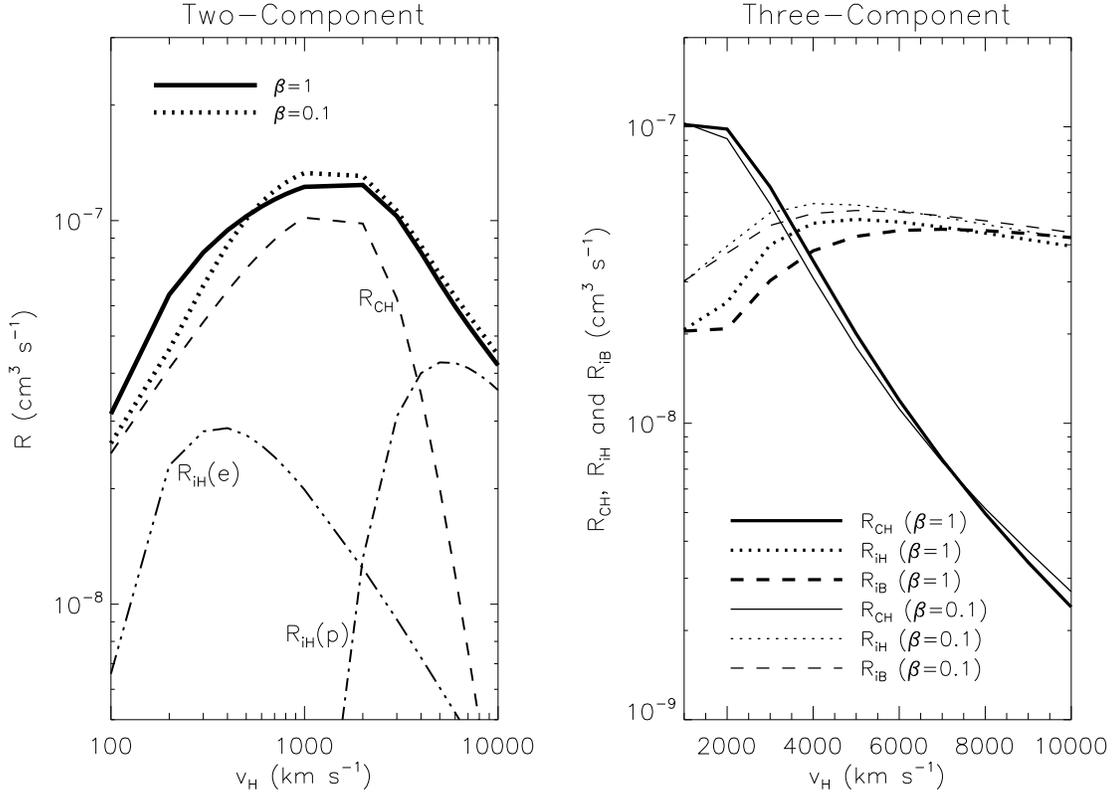}
\end{center}
\caption{Left: Rate coefficients for conversion of hydrogen atoms to protons ($p$) and electrons ($e$), used in the two-component model.  The velocity difference between the atoms and ions is $3v_{\rm{H}}/4$.  As an illustration, we display the individual rate coefficients for charge transfer ($R_{\rm{CH}}$; atoms and protons) and impact ionization ($R_{\rm{iH}}$; atoms, electrons and protons) for $\beta=1$, but only the total rate coefficient for $\beta=0.1$.  Right: Rate coefficients for interactions between atoms (H), ions (i) and broad neutrals (B), used in the three-component model.}
\label{fig:rates}
\end{figure}

We shall show how to account for the broad neutrals in \S\ref{sect:threefluid}.  In this section, we assume low neutral fractions ($f_{\rm{ion}} \gtrsim 0.9$) so that the broad neutrals will not affect the density of the ions significantly.  In this case, we can approximate the variation of mass, momentum and enthalpy flux with distance by
\begin{eqnarray}
&\frac{d}{dx}(n_{\rm{H}} v_{\rm{H}}) = - n_{\rm{H}} n R,\nonumber\\
&\frac{d}{dx}(n v) = n_{\rm{H}} n R,\nonumber\\
&\frac{d}{dx}\left(P + \rho v^2\right) = m_{\rm{H}} v_{\rm{H}} n_{\rm{H}} n R,\nonumber\\
&\frac{d}{dx}\left(Uv + Pv + \frac{1}{2}\rho v^3\right) = \frac{1}{2} m_{\rm{H}} v^2_{\rm{H}} n_{\rm{H}} n R,
\label{eq:set1}
\end{eqnarray}
where $\rho = m_p n$.  The internal energy density and pressure of the ion fluid are represented by $U$ and $P$, respectively.  The set of equations is supplemented by the following equations of state:
\begin{eqnarray}
&P = \left(\beta+1\right)nkT,\nonumber\\
&U = \frac{P}{\gamma-1} = \frac{3}{2} \left(\beta+1\right) nkT.
\end{eqnarray}
In the case of full equilibration ($\beta = 1$), we have $P = 2nkT$, where the factor of 2 accounts for contributions from both electrons and protons.  The polytropic index is $\gamma=5/3$ for a monoatomic gas.  The rate coefficient, $R$, characterizes the loss of atoms by conversion to ions via both charge transfer and ionization.  Examples of $R$ are shown in the left panel of Fig. \ref{fig:rates}; these rates assume a relative velocity of $3v_{\rm{H}}/4$ between the atoms and the peak of the ion distribution.   For $\beta=1$, charge transfer dominates over ionization at $v_{\rm{H}} \lesssim 4000$ km s$^{-1}$.  Details on how to calculate $R$ for a general relative velocity are given in Appendix \ref{append:rates}.

Suppose we can define a typical value for the rate coefficient, $ \bar{R}$.  Then a natural length scale for the problem emerges:
\begin{equation}
L = \frac{v_{\rm{H}}}{n_a \bar{R}},
\label{eq:lengthscale}
\end{equation}
where $n_a$ is the pre-shock atomic density.  Knowledge of $L$ allows us to define the dimensionless distance, $\zeta \equiv x/L$.  Other dimensionless variables follow naturally: $\eta \equiv n/n_a$, $\eta_{\rm{H}} \equiv n_{\rm{H}}/n_a$, $u \equiv v/v_{\rm{H}}$, $\epsilon \equiv kT/m_p v^2_{\rm{H}}$, and $\sR \equiv R/\bar{R}$.  The system of equations then becomes:
\begin{eqnarray}
&\frac{d\eta_{\rm{H}}}{d\zeta} = -\eta_{\rm{H}} \eta \sR,\nonumber\\
&\frac{d}{d\zeta}(\eta u) = \eta_{\rm{H}} \eta \sR,\nonumber\\
&\frac{d}{d\zeta}\left[ \eta u^2 + \left(\beta+1\right) \eta \epsilon \right] = \eta_{\rm{H}} \eta \sR,\nonumber\\
&\frac{d}{d\zeta}\left[ \eta u^3 + 5\left(\beta+1\right) \eta \epsilon u \right] = \eta_{\rm{H}} \eta \sR.
\label{eq:set2}
\end{eqnarray}

\subsection{GENERALIZED CONSERVATION EQUATIONS}
\label{subsect:conserve}

The upstream, initial values for $\eta_{\rm{H}}$, $\eta$, $u$ and $\epsilon$ are 1, $f_u=f_{\rm{ion}}/(1 - f_{\rm{ion}})$, 1 and $\epsilon_u \equiv kT_u/m_p v^2_{\rm{H}}$, respectively.  Adding the first pair of equations in (\ref{eq:set2}) and applying initial conditions, one obtains
\begin{equation}
\label{eq:etaconst}
\eta_{\rm{H}} + \eta u = 1 + f_u,
\end{equation}
which is a statement of the conservation of mass density flux, from which one can derive
\begin{equation}
\label{eq:etahDE}
\frac{d\eta_{\rm{H}}}{d\zeta} = -\frac{\eta_{\rm{H}} \sR}{u} \left( 1 + f_u - \eta_{\rm{H}} \right).
\end{equation}
Algebraic manipulation of the equations in (\ref{eq:set2}) yields the cubic equation for $u$:
\begin{equation}
\label{eq:ucubic}
(u-1) \left[ 4\eta u^2 - \eta u - 5\left(\beta+1\right) f_u \epsilon_u \right] = 0,
\end{equation}
With the solutions for $\eta_{\rm{H}}$, $\eta$ and $u$ in hand, one can then solve for $\epsilon$ using
\begin{equation}
\label{eq:epsilonconst}
\epsilon = \frac{f_u \epsilon_u}{\eta} + \frac{u}{\beta+1}\left( 1 - u \right).
\end{equation}

In general, since the equations in (\ref{eq:set2}) have the same source/sink term, we can add/subtract any given pair and obtain algebraic equations equated to six conservation constants.  Only three of the conservation constants are independent; the other three are a simple, linear combination of the first three.  One can choose to tackle the system of four coupled differential equations or substitute up to three of them by the conservation equations.

\subsection{ENERGY LOSSES FROM THE ATOMIC BEAM}

The atomic beam suffers energy losses due to excitation and ionization, prior to engaging in charge transfer events with the ions.  On average, an atom gets excited $N_{E} \sim 1.5$ times before it gets taken out of the beam (HM07).  Therefore, the final equation in (\ref{eq:set2}) has to be modified:
\begin{equation}
\frac{d}{d\zeta}\left[ \eta u^3 + 5\left(\beta+1\right) \eta \epsilon u \right] = \eta_{\rm{H}} \eta \sR \left(1 - 2 \epsilon_{\rm{loss}}\right).
\end{equation}
The quantity $\epsilon_{\rm{loss}} \equiv E_{\rm{loss}} / m_{\rm{H}} v^2_{\rm{H}}$ characterizes the energy loss,
\begin{equation}
E_{\rm{loss}} = 13.6 \mbox{ eV} + N_{E} \left(10.2 \mbox{ eV} \right).
\end{equation}
The cubic equation for $u$ changes slightly:
\begin{equation}
4\eta u^3 - 5\eta u^2 + u\left[\eta \left(1 - 2\epsilon_{\rm{loss}} \right) - 5 \left( \beta + 1 \right) f_u \epsilon_u \right] + 5 \left( \beta + 1 \right) f_u \epsilon_u + 2\epsilon_{\rm{loss}} = 0.
\end{equation}

We emphasize that this is an approximate way to account for the energy losses.  More generally, $N_{E}$ has a velocity dependence.  However, $\epsilon_{\rm{loss}} \sim 10^{-3}$ at $v_{\rm{H}} \sim 1000$ km s$^{-1}$, and a more complicated treatment (e.g., Cox \& Raymond 1985) is only important for shocks with $v_{\rm{H}} \lesssim 200$ km s$^{-1}$.  Excitation and subsequent emission of Ly$\alpha$ photons extract increments of 10.2 eV from the atomic beam; energy losses due to Balmer and other Lyman lines are a $\sim 10\%$ effect for $\sim 1000$ km s$^{-1}$ shocks.

\section{RESTRICTED THREE-COMPONENT MODEL}
\label{sect:threefluid}

At low and intermediate values of $f_{\rm{ion}}$, the creation of the broad neutrals competes effectively with that of the ions.  In this case we must generalize the hydrodynamical equations to a three-component model, consisting of the beam of atoms (``H''), the electrons and protons (``i''), and the broad neutrals (``B'').  Here we assume that the broad neutrals and ions share a common fluid velocity, $v_{\rm{B}} = v$, which implies that they share a common temperature as well ($T_{\rm{B}} = T$).  This approximation has been employed in all of the papers before HM07, who showed that the steep velocity dependence of the charge transfer cross section creates significant differences between $v_{\rm{B}}$ and $v$ (and hence $T_{\rm{B}}$ and $T$) for shock velocities $\gtrsim 3000$ km s$^{-1}$.  The full three-component model requires $v_{\rm{B}} \ne v$, which is beyond the scope of our present study (see \S\ref{sect:future}).

We consider three processes between the particles: charge transfer between the atoms and ions (with a rate coefficient of $R_{\rm{CH}}$), ionization of atoms by ions ($R_{\rm{iH}}$), and ionization of broad neutrals by ions ($R_{\rm{iB}}$); the rate coefficients are calculated using the formalism of HM07 and shown in the right panel of Fig. \ref{fig:rates}.  We neglect interactions between the atomic beam and the broad neutrals.

Under these assumptions, the equations in (\ref{eq:set1}) generalize to:
\begin{eqnarray}
&\frac{d}{dx} \left(n_{\rm{H}} v_{\rm{H}}\right) = -n_{\rm{H}} n(R_{\rm{CH}}+R_{\rm{iH}}),\nonumber\\
&\frac{d}{dx} \left(n_{\rm{B}} v\right) = n_{\rm{H}} n R_{\rm{CH}} - n_{\rm{B}} n R_{\rm{iB}},\nonumber\\
&\frac{d}{dx} \left(n v\right) = n_{\rm{H}} n R_{\rm{iH}}+n_{\rm{B}} n R_{\rm{iB}},\nonumber\\
&\frac{d}{dx} \left[\left(P + P_{\rm{B}} \right) + \frac{1}{2} m_p n v^2 + \frac{1}{2} m_{\rm{H}}n_{\rm{B}} v^2\right] = m_{\rm{H}} v_{\rm{H}} n_{\rm{H}} n \left(R_{\rm{CH}}+R_{\rm{iH}}\right),\nonumber\\
&\frac{d}{dx} \left[\left(U + U_{\rm{B}} \right)v + \left(P + P_{\rm{B}}\right)v + \frac{1}{2} m_p n v^3 + \frac{1}{2} m_{\rm{H}} n_{\rm{B}} v^3\right] = \frac{1}{2} m_{\rm{H}} v^2_{\rm{H}} n_{\rm{H}} n \left(R_{\rm{CH}}+R_{\rm{iH}}\right),
\end{eqnarray}
where $P$ and $U$ have the same definitions as before, while $P_{\rm{B}} = n_{\rm{B}} k T_{\rm{B}}$ and $U_{\rm{B}} = 3 P_{\rm{B}}/2$.

Casting the equations in dimensionless form yields:
\begin{eqnarray}
&\frac{d\eta_{\rm{H}}}{d\zeta} = -\eta_{\rm{H}} \eta \left(1+f_{\rm{iH}}\right)\sR,\nonumber\\
&\frac{d}{d\zeta} \left(\eta_{\rm{B}} u \right) = \eta_{\rm{H}} \eta \sR - \eta_{\rm{B}}\eta f_{\rm{iB}}\sR,\nonumber\\
&\frac{d}{d\zeta} \left(\eta u\right) = \eta_{\rm{H}} \eta f_{\rm{iH}}\sR + \eta_{\rm{B}} \eta f_{\rm{iB}}\sR,\nonumber\\
&\frac{d}{d\zeta} \left[ \left( \eta_t + \beta \eta \right) \epsilon + \eta_t u^2 \right] = \eta_{\rm{H}} \eta \left(1+f_{\rm{iH}}\right)\sR,\nonumber\\
&\frac{d}{d\zeta} \left[ 5\left(\eta_t + \beta \eta\right)\epsilon u + \eta_t u^3 \right] = \eta_{\rm{H}} \eta \left(1+f_{\rm{iH}}\right)\sR,
\label{eq:set3}
\end{eqnarray}
where $\eta_{\rm{H}}$, $\eta$, $u$ and $\epsilon$ retain their previous definitions, while $\eta_{\rm{B}} = n_{\rm{B}}/n_a$, $f_{ij} = R_{ij}/R_{\rm{CH}}$, $\sR = R_{\rm{CH}}/\bar{R}$ and $\eta_t \equiv \eta + \eta_{\rm{B}}$.  The initial conditions are: $\eta_{\rm{H}}(0)=1$, $\eta(0)=f_u$, $\eta_{\rm{B}}(0)=0$, $u(0)=1$ and $\epsilon(0)=\epsilon_u$.

Our approach to solving for the hydrodynamical variables is similar to the one previously described in \S\ref{subsect:conserve}.  In fact, the cubic equation for $u$ remains the same as equation (\ref{eq:ucubic}), but with $\eta$ replaced by $\eta_t$:
\begin{equation}
\label{eq:ucubic2}
(u-1) \left[ 4\eta_t u^2 - \eta_t u - 5\left(\beta+1\right) f_u \epsilon_u \right] = 0.
\end{equation}
Adding the first three equations in (\ref{eq:set3}) and applying initial conditions yields:
\begin{equation}
\label{eq:etaconst2}
\eta_{\rm{H}}+\eta_t u = 1+f_u.
\end{equation}
Defining $\mu \equiv \eta u$ and $\mu_{\rm{B}} \equiv \eta_{\rm{B}} u$, we use equation (\ref{eq:etaconst2}) to eliminate $\eta_{\rm{H}}$ and obtain:
\begin{eqnarray}
&\frac{d\mu}{d\zeta}  = \frac{1}{u} \left( 1+f_u \right) f_{\rm{iH}} \sR \mu - \frac{1}{u} f_{\rm{iH}} \sR \mu^2 - \frac{1}{u} \left( f_{\rm{iH}}- \frac{1}{u}f_{\rm{iB}}\right)\sR\mu_{\rm{B}} \mu,\nonumber\\
&\frac{d\mu_{\rm{B}}}{d\zeta} = \frac{1}{u} \left( 1+f_u \right) \sR \mu - \frac{1}{u} \sR \mu^2- \frac{1}{u} \left( 1+ \frac{1}{u} f_{\rm{iB}}\right)\sR\mu_{\rm{B}} \mu.
\label{eq:mu_Hmu_B}
\end{eqnarray}
Once $\eta$ and $\eta_{\rm{B}}$ are known, $\epsilon$ is determined using
\begin{equation}
\epsilon=\frac{\eta_tu \left(1-u\right)+ \left(1+\beta \right) f_u\epsilon_u}{\eta_t+\beta\eta}.
\label{eq:epsilonthreefluid}
\end{equation}

\section{SOLUTION METHODS}
\label{sect:solutions}

\subsection{APPROXIMATE SOLUTIONS AND THEIR ASYMPTOTES FOR THE TWO-COMPONENT MODEL}
\label{subsect:approx}

For $v_{\rm{H}} \gtrsim 1000$ km s$^{-1}$, we can assume $\epsilon_{\rm{loss}} \approx 0$ and constant $\sR$ (see \S\ref{sect:results}) behind the shock, and derive approximate solutions to the hydrodynamic variables and their corresponding asymptotes.  Again assuming upstream initial conditions, we obtain from equation (\ref{eq:ucubic}):
\begin{equation}
u = \frac{1}{8} \left[ 1 \pm \sqrt{1 + \frac{80\left( \beta + 1 \right) f_u \epsilon_u}{\eta} } \right].
\label{eq:u}
\end{equation}
We pick the positive root to ensure that $u > 0$.  Consider the limiting case where $f_u \epsilon_u \ll 1$.  We then have
\begin{equation}
u \approx \frac{1}{4} + \frac{5\left( \beta + 1 \right) f_u \epsilon_u}{\eta}.
\end{equation}
Using equation (\ref{eq:etaconst}), one gets
\begin{equation}
\eta \approx 4\left[ 1 + f_u - \eta_{\rm{H}} - 5\left( \beta + 1 \right) f_u \epsilon_u \right].
\end{equation}
Substituting this into the first equation in (\ref{eq:set2}) yields the Bernoulli equation, $\eta^\prime_{\rm{H}} + a_u \eta_{\rm{H}} = b_u \eta^2_{\rm{H}}$, where $a_u = 4 \sR [1 + f_u - 5(\beta+1) f_u \epsilon_u]$ and $b_u = 4 \sR$.  Solving for $\eta_{\rm{H}}$ yields:
\begin{equation}
\eta_{\rm{H}} \approx \left[ \frac{b_u}{a_u} + \left(1 - \frac{b_u}{a_u}\right) \exp{\left( a_u\zeta \right)} \right]^{-1}.
\end{equation}

The set of solutions, $(u,\eta,\eta_{\rm{H}},\epsilon)$, has the following asymptotes ($\zeta \gg 1$):
\begin{eqnarray}
&u \rightarrow \frac{1}{4},\nonumber\\
&\eta \rightarrow 4,\nonumber\\
&\eta_{\rm{H}} \rightarrow 0,\nonumber\\
&\epsilon \rightarrow \frac{3}{16(\beta+1)},
\end{eqnarray}
consistent with standard jump conditions for a strong shock.  Another quantity of interest is the Mach number, $M$, of the post-shock ion flow.  Since $P \propto \rho^{5/3}$, we have $c^2_s = \partial P/ \partial \rho = 5(\beta + 1)\epsilon v^2_{\rm{H}}/3$.
The Mach number then becomes
\begin{equation}
M = \frac{v}{c_s} = u ~\sqrt{\frac{3}{5\left( \beta + 1 \right)\epsilon}} \rightarrow \frac{1}{\sqrt{5}}.
\end{equation}

\subsection{NUMERICAL SOLUTIONS}
\label{subsect:numerical}

To obtain numerical solutions for the set of equations in (\ref{eq:set2}), we employ a coordinate system where the beginning of the transition zone (Fig. \ref{fig:zone}) is placed at $\zeta = 0$.  We discuss our approach for $\epsilon_{\rm{loss}}=0$; our solution method for $\epsilon_{\rm{loss}} \neq 0$ is conceptually similar.  Firstly, we assume $\eta_{\rm{H}}(0)=1$, and use equations (\ref{eq:etaconst}), (\ref{eq:ucubic}), and (\ref{eq:epsilonconst}) to solve for $\eta(0)$, $u(0)$, and $\epsilon (0)$.  To obtain a self-consistent set of solutions, we use an iterative approach.  For the first iteration, the (dimensionless) ion velocity and temperature are taken to be $u^{(1)}(\zeta)=u(0)$ and $\epsilon^{(1)}(\zeta)=\epsilon(0)$, respectively.  Equation (\ref{eq:etahDE}) is then solved using a standard Runge-Kutta method for $\eta_{\rm{H}}^{(1)}(\zeta)$, which is substituted into equation (\ref{eq:etaconst}) to determine $\eta^{(1)}(\zeta)$.  With $\eta^{(1)}(\zeta)$ known, the cubic equation in (\ref{eq:ucubic}) is solved at each $\zeta$ using a simple bisection method in the range $0<u<1$, for which there is only one physical root.  (Note that if $\epsilon_{\rm loss} \ll1$, we only need to solve a quadratic equation for $u^{(2)}(\zeta)$.)  The updated value of $u^{(2)}(\zeta)$ is substituted into equation (\ref{eq:epsilonconst}), yielding $\epsilon^{(2)}(\zeta)$.  This procedure is iterated $i$ times, until the dependent variables $\eta^{(i)}_{\rm{H}}(\zeta)$, $\eta^{(i)}(\zeta)$, $u^{(i)}(\zeta)$ and $\epsilon^{(i)}(\zeta)$ converge.  The convergence is monitored in two ways: 1. The final, fractional correction for each dependent variable must be less than a pre-determined tolerance, i.e., $\vert u^{(i)}(\zeta)-u^{(i-1)}(\zeta)\vert/u^{(i)}(\zeta) \la \epsilon_{\rm tol}$; 2. The mass, momentum, and enthalpy fluxes must be constant and equal to their values at $\zeta=0$.

In practice, the values of the variables converge rapidly during the iteration, requiring $i < 10$.  For shock velocities $v_s \ga 1000$ km s$^{-1}$, the velocity difference between the atomic and ion populations is nearly constant, meaning $R\approx \bar{R}$ ($\sR \approx 1$) throughout the shock transition zone.  In this regime, $\epsilon_{\rm loss}$ is negligible and the analytical solutions of \S\ref{subsect:approx} are excellent approximations to the two-component numerical calculations.  At lower shock velocities ($v_{\rm{H}} \lesssim 200$ km s$^{-1}$), variations in $u(\zeta)$ and non-negligible values of $\epsilon_{\rm loss}$ quantitatively change the solution, requiring the full numerical treatment.

The numerical method in the three-component case is analogous to the iterative procedure described above.  Assuming $\eta_{\rm{H}}(0)=1$ and $\eta_{\rm{B}}(0)=0$, equations (\ref{eq:ucubic2}), (\ref{eq:etaconst2}), and (\ref{eq:epsilonthreefluid}) are solved for $\eta(0)$, $u(0)$, and $\epsilon(0)$.  As before, we set $u^{(1)}=u(0)$, and now solve the equations in (\ref{eq:mu_Hmu_B}) for $\mu^{(1)}$ and $\mu_{\rm{B}}^{(1)}$ using a standard Runge-Kutta algorithm, from which we calculate $\eta^{(1)}$ and $\eta^{(1)}_{\rm{B}}$ using $u^{(1)}$.  Equation (\ref{eq:ucubic2}) is solved using the updated values for the (dimensionless) densities to give the improved estimate, $u^{(2)}$.  The process is then iterated in the manner described above.

\section{RESULTS}
\label{sect:results}

\begin{figure}
\begin{center}
\includegraphics[width=6in]{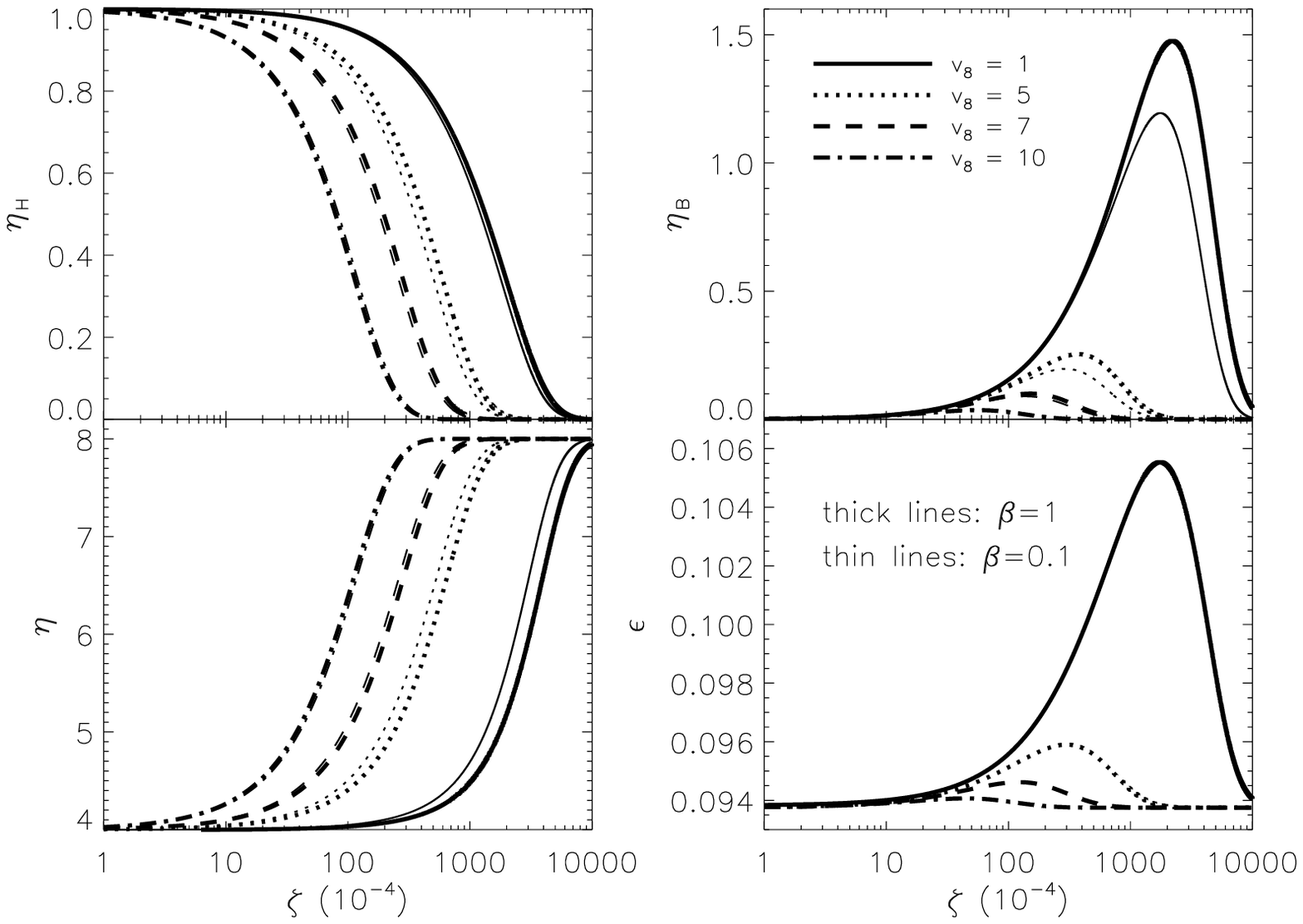}
\end{center}
\caption{Spatial variation of $\eta_{\rm{H}}$, $\eta$, $\eta_{\rm{B}}$ and $\epsilon$ in the three-component model, for various values of the shock velocity, $v_8 = v_{\rm{H}}/1000$ km s$^{-1}$, and $f_{\rm{ion}}=0.5$.  Like in the two-component model, $u \approx 1/4$ throughout the shock transition zone.}
\label{fig:structure}
\end{figure}

Figure \ref{fig:structure} displays the structure of the shock transition zone for $v_8 = v_{\rm{H}}/1000$ km s$^{-1}$ = 1, 5, 7 and 10.  In each case, we assume $f_u = 1$ ($f_{\rm{ion}}=0.5$) and $T_4 = T_u/10^4$ K = 1.  Knowledge of $v_8$ and $T_4$ then determines $\epsilon_u = 8.25 \times 10^{-5} ~T_4/v^2_8$.  The pre-shock atomic density is $n_a =n_0/(1+f_u) \sim 0.1$ cm$^{-3}$.  For example, in the case of SN 1006, R07 find $0.25 \le n_0 \le 0.4$ cm$^{-3}$, implying $0.025 \le n_a \le 0.04$ cm$^{-3}$.

Hydrodynamical quantities vary from their pre- to post-shock values, at the beginning of the shock transition zone, according to the Rankine-Hugoniot jump conditions.   Across the zone, the ions have no velocity structure, consistent with the assumption made by CKR80.  Numerically, for a strong shock, $u$ goes from 1 to 1/4 immediately and $\eta$ jumps to $4f_u$ to conserve ion mass flux.  The latter is in accordance with the Rankine-Hugoniot density jump (Zel'dovich \& Raizer 1966) of
\begin{equation}
J_{d_0} = \frac{\left(\gamma+1\right)\epsilon_{d_0} + \left(\gamma-1\right)\epsilon_u}{\left(\gamma-1\right)\epsilon_{d_0} + \left(\gamma+1\right)\epsilon_u},
\label{eq:jump}
\end{equation}
where $\epsilon_{d_0}$ is the value of $\epsilon$ immediately after the shock (and not far downstream).  For $\epsilon_{d_0} \gg \epsilon_u$ and $\gamma = 5/3$, we recover the familiar density jump of 4 for a monoatomic gas.  For strong shocks ($v_{\rm{H}} \gtrsim 1000$ km s$^{-1}$), energy losses from the atomic beam are negligible after the jump, and the downstream density eventually evolves to $J_d \approx 4(1+f_u) \approx 8$ for $f_u = 1$.  A factor of 4 comes from the jump condition, while an additional factor of 2 results from adding the atoms to the population of ions.  We note that for weak shocks, a departure from the decrease by a factor of 4 in $u$ occurs, consistent with equation (\ref{eq:jump}).  The departure from a jump of 4 in $\eta$ (and its subsequent evolution) follows naturally to conserve momentum; the asymptotic value of $\epsilon$ dips to below 3/32 ($\beta=1$) due to energy losses.  It is worth noting that as the shock velocity increases, the distinction between $\eta_{\rm{B}}$ for $\beta=0.1$ and 1 vanishes.

With a telescope having sufficient angular resolution, it may be possible to measure the spatial emissivity profiles of the narrow and broad H$\alpha$ lines.  The narrow emissivity is given by $\xi_n = n n_{\rm{H}} R_{\rm{H}\alpha,n}$, where $R_{\rm{H}\alpha,n}$ is the rate coefficient for the narrow H$\alpha$ line (cf. equation [20] in HM07).  Broad emission has two distinct contributions: (1) from charge transfers of the atoms in the initial beam directly to excited states of the broad neutrals, with a rate coefficient given by $R_{\rm{H}\alpha,b_0}$; and (2) from excitations of and charge transfers between the broad neutrals to their excited states, with a net rate coefficient of $R_{\rm{H}\alpha,b_*}$.  The addition of $R_{\rm{H}\alpha,b_0}$ and $R_{\rm{H}\alpha,b_*}$ yields the broad-line rate coefficient (cf. equation [22] in HM07).  Charge transfers involving atoms from the beam will naturally have the spatial profile of the narrow emissivity.

\begin{figure}
\begin{center}
\includegraphics[width=6in]{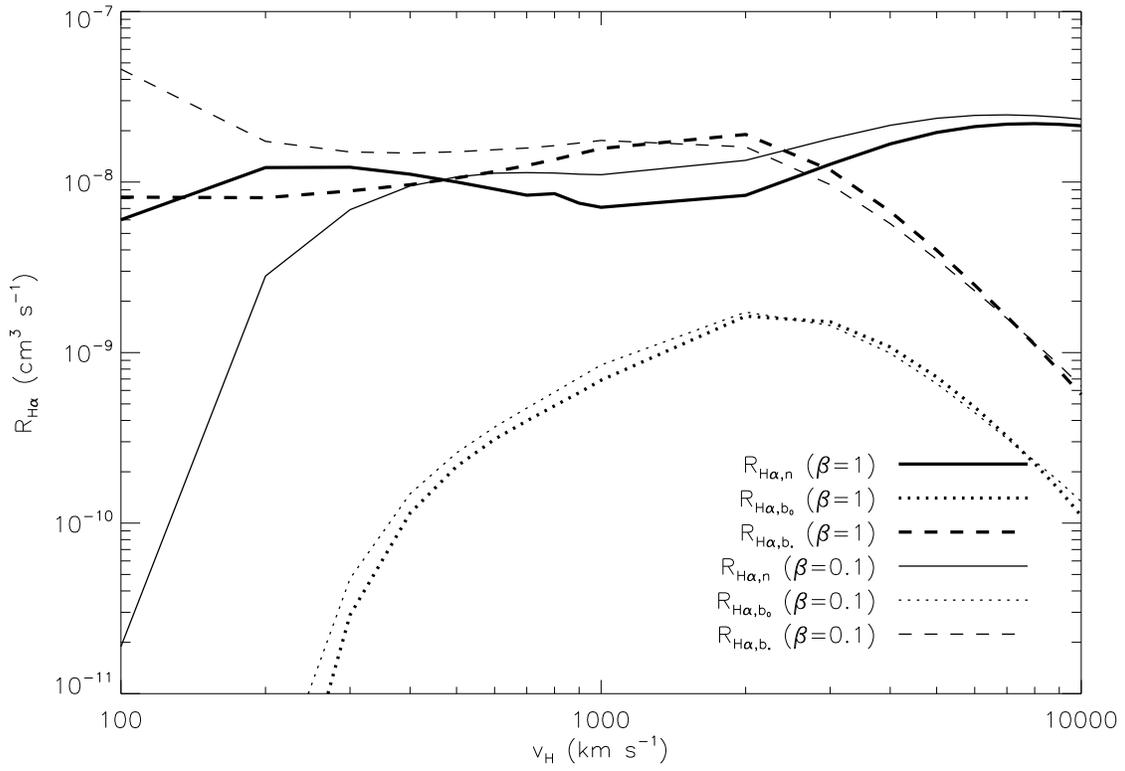}
\end{center}
\caption{Rate coefficients for the narrow ($R_{\rm{H}\alpha,n}$) and broad ($R_{\rm{H}\alpha,b_0} + R_{\rm{H}\alpha,b_*}$) H$\alpha$ lines, assuming Case B and A conditions, respectively.}
\label{fig:rates2}
\end{figure}

We first derive the emissivity profile of the broad line, $\xi_b$, from $\xi_n$, using the two-component model: upon its creation at $x_0$, a broad neutral drifts for an average distance of $l_d = l_d(x,x_0)$ until it gets destroyed by impact ionization.  The drift velocity, $v_d$, is the velocity difference between the peak of the broad neutral distribution and the shock front, $1/4 \la v_d/v_{\rm{H}} \la 1$.
The broad H$\alpha$ emissivity is then
\begin{equation}
\xi_b(x) = \frac{R_{\rm{H}\alpha,b_0}}{R_{\rm{H}\alpha,n}} \xi_n + \frac{R_{\rm{H}\alpha,b_*}}{R_{\rm{H}\alpha,n}}\int^x_0 ~\xi_n(x^\prime) ~P\left(x,x^\prime \right) ~dx^\prime,\label{eq:broadxi}
\end{equation}
where $P(x,x_0) = P_0(x_0) \exp{[-(x-x_0)/l_d(x,x_0)]}$ is the ``transfer function''.  Details regarding $l_d(x,x_0)$ and $P(x,x_0)$ are given in Appendix \ref{append:transfer}.  Examples of $R_{\rm{H}\alpha,n}$, $R_{\rm{H}\alpha,b_0}$ and $R_{\rm{H}\alpha,b_*}$ are given in Fig. \ref{fig:rates2}.    In the three-component model, the broad and narrow H$\alpha$ emissivities are simply given by $\xi_b = n n_{\rm{H}} R_{\rm{H}\alpha,b_0} + n n_{\rm{B}} R_{\rm{H}\alpha,b_*}$ and $\xi_n = n n_{\rm{H}} R_{\rm{H}\alpha,n}$, respectively.

In the two-component model, the ratio of broad to narrow H$\alpha$ emission is given by
\begin{equation}
I_b/I_n = \frac{\int^\infty_0 \xi_b(x) ~dx}{\int^\infty_0 \xi_n(x) ~dx} = \frac{R_{\rm{H}\alpha,b_0} + R_{\rm{H}\alpha,b_*}}{R_{\rm{H}\alpha,n}},
\end{equation}
such it is equal to the ratio of rate coefficients and the integrated line ratio is preserved.  In the three-component model, this is not necessarily the case, as
\begin{equation}
L_{bn} = \frac{R_{\rm{H}\alpha,b_0} ~\int^\infty_0 n n_{\rm{H}} ~dx + R_{\rm{H}\alpha,b_*} ~\int^\infty_0 n n_{\rm{B}} ~dx}{R_{\rm{H}\alpha,n}~\int^\infty_0 n n_{\rm{H}} ~dx} \neq I_b/I_n,
\end{equation}
due to the fact that $\int^\infty_0 n n_{\rm{H}} ~dx \neq \int^\infty_0 n n_{\rm{B}} ~dx$ in general.

\begin{figure}
\begin{center}
\includegraphics[width=6in]{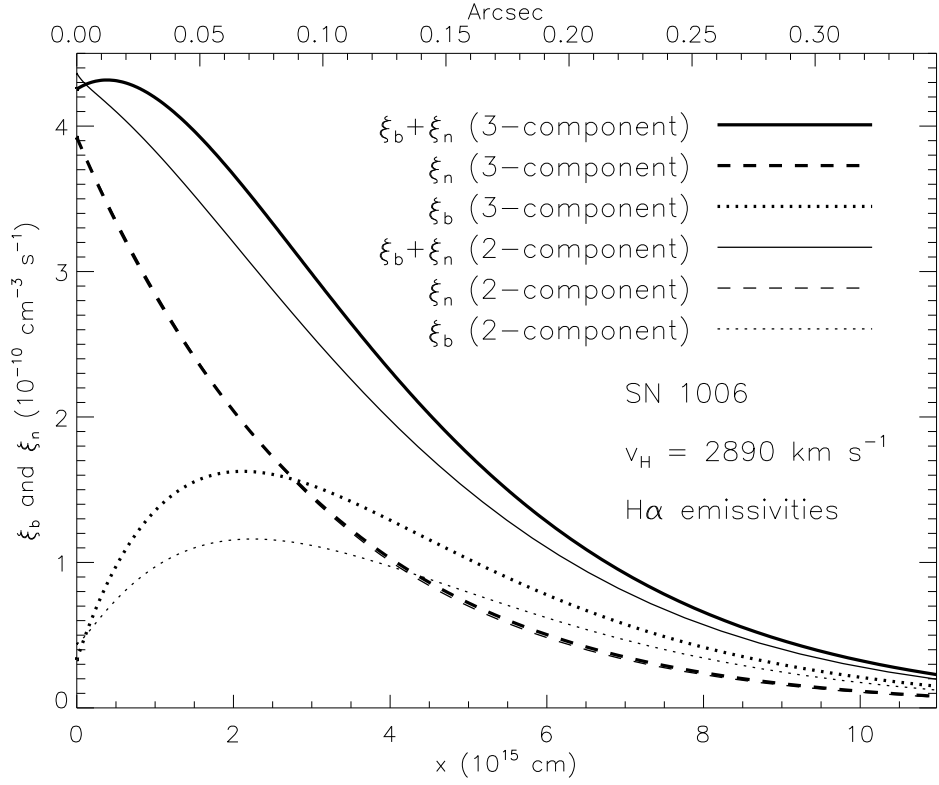}
\end{center}
\caption{Emissivity of the broad and narrow H$\alpha$ lines from a shock with parameters representative of SN 1006 (see text).}
\label{fig:sn1006}
\end{figure}

The high shock velocity and low neutral fraction of the remnant of supernova (SN) 1006, located 2.1 kpc away, makes it an ideal case study for both the two- and three-component models.  Following R07, we compute $\xi_b$ and $\xi_n$ (Fig. \ref{fig:sn1006}).  We adopt the following parameters: $v_{\rm{H}} =  2890$ km s$^{-1}$ (R07), $v_d/v_{\rm{H}} = 0.34$, $\beta=0.1$ (Ghavamian et al. 2002, hereafter G02), $f_{\rm{ion}} = f_u/(1+f_u) = 0.9$ (G02), $T_4 = 1$, $n_a$ = 0.025 cm$^{-3}$ ($n_0$ = 0.25 cm$^{-3}$; R07), $\bar{R} = 1.1 \times 10^{-7}$ cm$^3$ s$^{-1}$, $R_{\rm{CH}} = 5.9 \times 10^{-8}$ cm$^3$ s$^{-1}$, $R_{\rm{iH}} = 5.0 \times 10^{-8}$ cm$^3$ s$^{-1}$ and $R_{\rm{iB}} = 4.5 \times 10^{-8}$ cm$^3$ s$^{-1}$.  The broad H$\alpha$ line has the following parameters: $R_{\rm{H}\alpha,b_0} = 1.5 \times 10^{-9}$  cm$^3$ s$^{-1}$ and $R_{\rm{H}\alpha,b_*} = 1.0 \times 10^{-8}$  cm$^3$ s$^{-1}$ (Case A conditions).  The narrow H$\alpha$ line has: $R_{\rm{H}\alpha,n} = 6.9 \times 10^{-9}$  cm$^3$ s$^{-1}$ (Case A) and $1.7 \times 10^{-8}$  cm$^3$ s$^{-1}$ (Case B).  We emphasize that it is not our intention to model the Lyman line trapping, as done by Ghavamian et al. (2001) and G02.  Rather, we wish to calculate the relative shifts between $\xi_b$ and $\xi_n$, and demonstrate that $L_{bn} \ne I_b/I_n$; these are not dependent on the opacity assumptions for the narrow H$\alpha$ line.  As a matter of illustration, we adopt Case B conditions for the narrow H$\alpha$ line.

Except for $T_4$, values of the parameters without references were computed using the formalism of HM07.  The drift velocity is $v_d/v_{\rm{H}} > 1/4$ because at the velocity of SN 1006, charge transfer is not efficient enough to create a Maxwellian population of broad neutrals centered at $v_{\rm{H}}/4$.

Our calculation shows that $\xi_b$ peaks at $\sim 0\farcs$07 from the shock front, a factor of $\sim$ 2 smaller than the $\sim$ 0$\farcs$14 value computed by R07.  The smaller spatial scale of the current calculation is due to a numerical error in the model code used in R07; it implies that the lower values of $n_0$ in the curved shock models of Fig. 5 of R07 will not produce too much H$\alpha$ emission towards the inside of the remnant, and the density range $0.25 \le n_0 \le 0.4$ cm$^{-3}$ obtained by R07 produces too steep a falloff.  However, smaller pre-shock densities require larger radii of curvature to match the observed peak surface brightness, and that produces too gradual a falloff towards the outside of the remnant.  Overall, the revised models of R07 are compatible with densities in the range $0.15 \le n_0 \le 0.3$ cm$^{-3}$, but with small-scale ripples in the shock front that broaden the filament by about 0$\farcs$5.  The values of our emissivities are comparable to those of R07; minor discrepancies in the emissivities may be partially due to our use of cascade matrices to compute the H$\alpha$ rate coefficients, following HM07.  Excitation from the ground state tends to populate the lower $l$ levels, especially the $p$ ones.  Hence, our calculations over-estimate the cascade contribution, while R07 ignores it, and the true emissivities are probably bracketed by these two approaches.

For SN 1006, we compute $I_b/I_n = 0.68$ and $L_{bn} = 0.77$ (Case B), with the latter being about 13\% higher than the former.  Despite this difference, it is worthwhile to note that the shift between the peaks of $\xi_b$ and $\xi_n$ is about the same in both the two- and three-component calculations.  Furthermore, Smith et al. (1991) and G02 measure the broad-to-narrow H$\alpha$ line ratio to be about 0.73 and 0.84, respectively, and thus our prediction is within the range of uncertainty.

\section{DISCUSSION}
\label{sect:discussion}

\subsection{SNR 1987A \& BALMER-DOMINATED SUPERNOVA REMNANTS}
\label{subsect:bdsnr}

Balmer-dominated SNRs are named for the dominance of their hydrogen lines (over forbidden ones), as first described by Chevalier \& Raymond (1978) and CKR80.  They are characterized by their two-component spectra, which consists of a narrow ($\sim 10$ km s$^{-1}$) superimposed upon a broad ($\sim 1000$ km s$^{-1}$) line.  Broad line emission is produced when the atoms engage in charge transfer reactions with the post-shock ions.  Narrow lines are the result of direct excitation of the pre-shock atoms.  For a given value of $\beta$, the full width at half-maximum (FWHM) of the broad component is uniquely related to the shock velocity, $v_s=v_{\rm{H}}$, a relation which provides a way of measuring the distances to SNRs (Kirshner, Winkler \& Chevalier 1987).

As mentioned in \S\ref{sect:threefluid}, HM07 showed that the charge transfer cross section is sensitive to the shock velocity. For very fast shocks, the bulk velocity of the broad neutrals exceeds that of the protons, resulting in lower values of the FWHM for the broad neutral distribution relative to the one for the protons.  HM07 further suggested that the FWHM versus $v_s$ relation might be modified substantially if one takes the structure of the shock transition zone into account.  If the velocity difference between the atoms and the protons were considerably less than the $3v_s/4$ assumed in earlier models, then two consequences would result: 1. The amount of broad emission produced would be under-estimated; 2. For a given FWHM, the shock velocity inferred would always be {\it less} than the true value.  However, as we saw in \S\ref{sect:results}, the velocity of the proton fluid is decelerated to $v_{\rm{H}}/4$ almost immediately, and the subsequent evolution of $v$ is weak.  This validates the thin shock assumption made by CKR80 and HM07 and highlights a puzzle --- how does one account for the excessive amount of broad (``interior'') emission observed in SNR 1987A (Heng et al. 2006; HM07)?  More optical/ultraviolet spectroscopic studies of SNR 1987A are needed to shed light on the issue.  

\begin{figure}
\begin{center}
\includegraphics[width=6in]{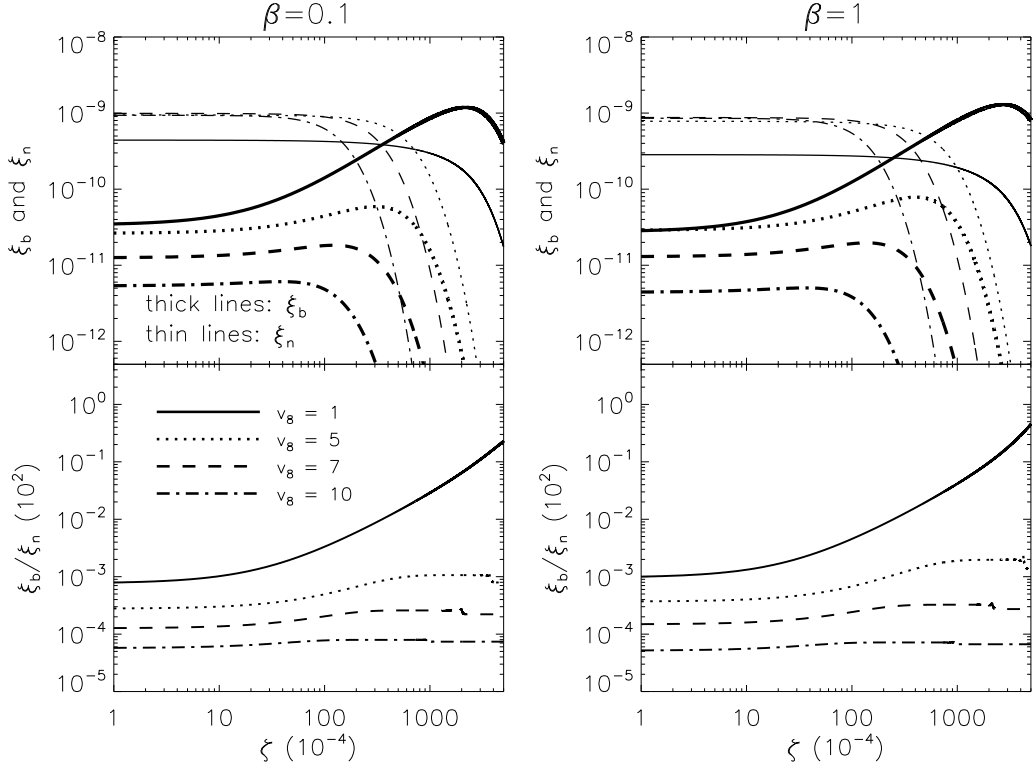}
\end{center}
\caption{Emissivity of the broad and narrow H$\alpha$ lines and their ratios, for various values of the shock velocity, $v_8 = v_{\rm{H}}/1000$ km s$^{-1}$, $f_{\rm{ion}}=0.5$ and $n_a = 0.1$ cm$^{-3}$.}
\label{fig:profiles}
\end{figure}

\begin{figure}
\begin{center}
\includegraphics[width=6in]{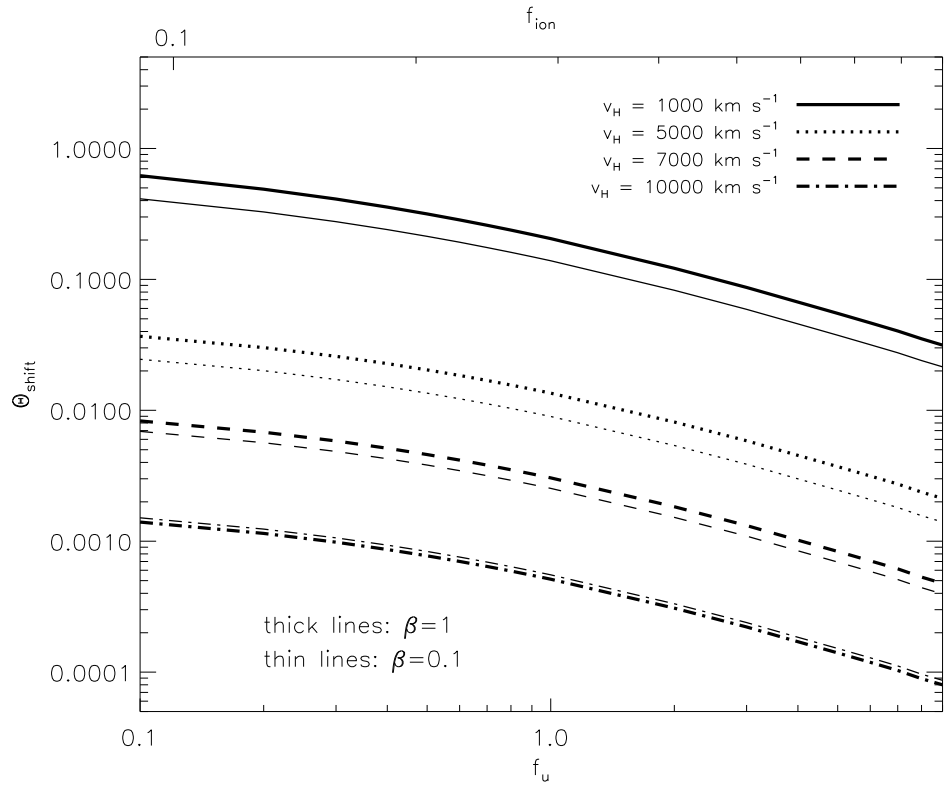}
\end{center}
\caption{Dimensionless spatial shift, $\Theta_{\rm{shift}}$, between the centroids of the broad and narrow H$\alpha$ line emissivity profiles.}
\label{fig:scaling}
\end{figure}

Another relevant quantity to examine is the spatial shift between the centroids\footnote{For an arbitrary distribution $F(x)$, the centroid is defined as the point $x_c$ such that $\int^{x_c}_{-\infty} F(x) ~dx = \int^\infty_{x_c} F(x) ~dx$.} of $\xi_b$ and $\xi_n$.  In Fig. \ref{fig:profiles}, we compute $\xi_b$ and $\xi_n$ for $f_{\rm{ion}}=0.5$ and $n_a = 0.1$ cm$^{-3}$, over the range $1000 \le v_{\rm{H}} \le 10,000$ km s$^{-1}$, as well as for $\beta=0.1$ and 1.  (Again, as a matter of illustration, we assume Case B conditions for the narrow H$\alpha$ line.)  Then, for $0.1 \lesssim f_{\rm{ion}} \lesssim 0.9$, we determine the {\it dimensionless} spatial shift, $\Theta_{\rm{shift}}$ (Fig. \ref{fig:scaling}), which is the spatial shift normalized by $L$.  At any given shock velocity, $\Theta_{\rm{shift}}$ decreases with increasing $f_{\rm{ion}}$.  This is explained by the fact that at high neutral fractions (low $f_{\rm{ion}}$), there are initially only a small number of ions available; the system drifts along until there are enough ions to create the broad neutrals.  Hence, the length scale for the creation of broad neutrals is relatively larger, corresponding to a greater shift in the centroid of $\xi_b$.  For a fixed value of $f_{\rm{ion}}$, $\Theta_{\rm{shift}}$ decreases as the shock velocity increases, for $v_{\rm{H}} \ge 1000$ km s$^{-1}$.  This is because charge transfer reactions are favored at lower velocities and the larger number of broad neutrals created ensures that the centroid of $\xi_b$ is shifted farther downstream.  This behavior is not true for shocks with $v_{\rm{H}} \lesssim 1000$ km s$^{-1}$, as the charge transfer cross section becomes nearly constant with velocity.

Observationally, it should be possible to measure $\Theta_{\rm{shift}}$ with WFC3 onboard the {\it Hubble Space Telescope} if one isolates the narrow H$\alpha$ component with a narrow band filter.  Such measurements can be used to constrain $n_0$ in some SNRs.

\subsection{FUTURE WORK}
\label{sect:future}

Though the velocity of the broad neutrals and ions has no spatial structure, the densities of the atoms, broad neutrals and ions vary across the width of the shock transition zone.  As mentioned in \S\ref{sect:intro}, knowledge of the density structures is relevant to modeling Ly$\alpha$ resonant scattering in young, core-collapse SNRs.  Photons are produced within the transition zone and resonantly scatter with a path length $l_{\rm{mfp}} \sim 10^{13} \mbox{ cm } t_{\rm{10yr}} ~T^{1/2}_{\rm{ej},100} < l_{\rm{zone}}$, where $t_{\rm{10yr}}$ is the time since the supernova explosion in units of 10 years and $T_{\rm{ej}} = 100 \mbox{ K } T_{\rm{ej},100}$ is the temperature of the freely streaming ejecta.  There is evidence for Ly$\alpha$ resonant scattering in SNR 1987A (Michael et al. 2003; Heng et al. 2006); while we now know how the density evolves spatially, it is beyond the scope of this paper to model the scattering process and observed spectra.  It is, however, worthy to note that for $v_s \gtrsim 1000$ km s$^{-1}$ ( $\epsilon_{\rm{loss}} \ll 1$) and $f_{\rm{ion}} \gtrsim 0.9$, the approximate, two-component solution for $n_{\rm{H}} = \eta_{\rm{H}} n_a$ becomes a good one.  The interested reader is referred to Zheng \&  Miralda-Escud\'{e} (2002) and Tasitsiomi (2006) for the physics of Ly$\alpha$ resonant scattering.

We have constructed a three-component model where we employed the simplifying assumption $v_{\rm{B}}=v$.  At $v_{\rm{H}} \gtrsim 3000$ km s$^{-1}$, impact excitation and ionization become competitive with charge transfer, and the resulting skewed Maxwellian of the broad neutrals peaks at $v_{\rm{B}} \ge v$.  The quantity $\int^\infty_0 n_{\rm{B}} ~dx$ is sensitive to changes in $v_{\rm{B}}$, which is relevant to the determination of $L_{bn}$.  Therefore, a $v_{\rm{B}} \neq v$ treatment is necessary (M. van Adelsberg et al. 2007, in preparation).  Before (CKR80 and HM07), the ratio of broad to narrow H$\alpha$ {\it rates}\footnote{In this paper, we define the emissivities as functions of $x$, i.e., $\xi(x)$.  The rate is defined as the value of $\xi(x)$ at a fixed value of $x$.} was assumed to be equal to the ratio of {\it rate coefficients}, as the densities of the atoms, ions and broad neutrals were assumed to be constant.  Physically, a full three-component model will answer the following question: for two Balmer-dominated SNRs with the same shock velocity, do we expect the ratio of broad to narrow H$\alpha$ {\it rates} to be the same if one is highly ionized and the other is largely neutral?

There are several important aspects of the hydrogen emission from non-radiative shocks that remain to be explored.  If a significant fraction of the energy dissipated by a shock goes into cosmic rays, a precursor will heat and accelerate the gas before it reaches the shock transition zone, altering the density and velocity jumps and changing the post-shock ion temperatures (Blandford \& Eichler 1987; Drury et al. 2001).  Evidence for such a precursor is found in the anomalously large widths of the narrow H$\alpha$ lines in Balmer-dominated shocks (Sollerman et al. 2003).  Furthermore, when a pre-shock atom is ionized upstream, it effectively becomes a ``pick-up ion'', analogous to those observed in inter-planetary space (Kallenbach et al. 2000).  These pick-up ions are preferentially accelerated to become anomalous cosmic rays.  In SNR shocks, they may form an isotropic, mono-energetic population that might perturb the Balmer line profiles.

\acknowledgments \scriptsize
K.H. is grateful to Roger Chevalier, Jeremy Darling, Bruce Draine, Claes Fransson, Peter Goldreich, Robert Kirshner, Davide Lazzati, Peter Lundqvist, Rosalba Perna, Jeffrey Weiss and Jared Workman for illuminating conversations.  We thank the anonymous referee for his/her meticulous reading and critical comments which substantially improved the manuscript.  K.H. acknowledges Chandra Grant XXX, and thanks both the Max Planck Institutes for Astrophysics (MPA) and Extraterrestrial Physics (MPE) for their generous hospitality during the summer and fall of 2007; he is eagerly anticipating a future position at the Institute for Advanced Study (Princeton).  M.v.A. acknowledges NSF Grant XXX.

\normalsize


\appendix
\section{APPENDIX: RATE COEFFICIENT}
\label{append:rates}

The reaction rate coefficient between a beam of atoms and the Maxwellian population of an ion species $s$ ($p$ for protons and $e$ for electrons) takes the form:
\begin{eqnarray}
R_s(v,T;v_{\rm{H}},\sigma,m_s) &= \int \int f_{\rm{M}}\left(\vec{v}_1,T;m_s \right) ~\delta\left(\vec{v}_2 - \vec{v}_{\rm{H}} + \vec{v} \right) ~\sigma\left(\vert \vec{v}_1 - \vec{v}_2 \vert\right) ~\vert \vec{v}_1 - \vec{v}_2 \vert ~d^3v_1 ~d^3v_2\nonumber\\
&= 2 \pi ~f_{\rm{M},0} \int^\infty_{-\infty} \int^\infty_0 ~\exp{\left[-\frac{m_s\left(v^2_r + v^2_z \right)}{2kT}\right]} ~\sigma\left(\Delta v\right) ~\Delta v ~v_r ~dv_r ~dv_z,
\end{eqnarray}
where $d^3v_1 = 2\pi v_r dv_r dv_z$ and $\Delta v = \sqrt{ v^2_r + \left(v_z - \vert v_{\rm{H}} - v \vert \right)^2}$.  The velocity difference between the atoms and the centroid of the ions is $ \vert v_{\rm{H}} - v \vert$, which is $3v_{\rm{H}}/4$ in the models of CKR80 and HM07.   The coefficient in front of the Maxwellian is $f_{\rm{M},0} = (m_s/2\pi k T)^{3/2}$.

The rate coefficient summed over species is
\begin{equation}
R = R_p(v,T;v_{\rm{H}},\sigma_{I,p} + \sigma_{T,p},m_p) + R_e(v,\beta T;v_{\rm{H}},\sigma_{I,e},m_e),
\end{equation}
since the atoms and protons participate in both ionization (``$I$'') and charge transfer (``$T$'') events, while the atoms and electrons only interact via the former process (if one neglects impact excitation).  With the exception of that for charge transfer to an excited state (Barnett 1990), all of the cross sections are taken from Janev \& Smith (1993).

\section{APPENDIX: TRANSFER FUNCTION \& DRIFT LENGTH}
\label{append:transfer}

The probability for destroying a broad neutral via impact ionization is described by
\begin{equation}
\frac{dP}{dx} = - \frac{n R_{\rm{iB}}}{v_d} P.
\end{equation}
Upon its creation at $x_0$, a broad neutral has a $P_0=P_0(x_0)$ chance of survival.  The probability of it drifting to a location $x$ is given by integrating the previous equation from $x_0$ to $x$:
\begin{equation}
P(x,x_0) = P_0(x_0) \left[\frac{C_1 + C_2 \exp{\left(C_3 x_0\right)}}{C_1 + C_2 \exp{\left(C_3 x\right)}}\right]^\alpha,
\end{equation}
where $\alpha \equiv R_{\rm{iB}} v_{\rm{H}} / \bar{R} v_d$, $C_1 \equiv b_u/a_u$, $C_2 \equiv 1 - C_1$ and $C_3 \equiv a_u/L$.  Since the mass flux of broad neutrals has to be conserved, we require
\begin{equation}
\int^\infty_{x_0} ~P(x,x_0) ~dx = 1.
\end{equation}
If we solve for $P_0$ analytically, it contains the hypergeometric function of Gauss, $_2F_1$; we choose instead to seek numerical solutions of $P_0$.

\begin{figure}
\begin{center}
\includegraphics[width=6in]{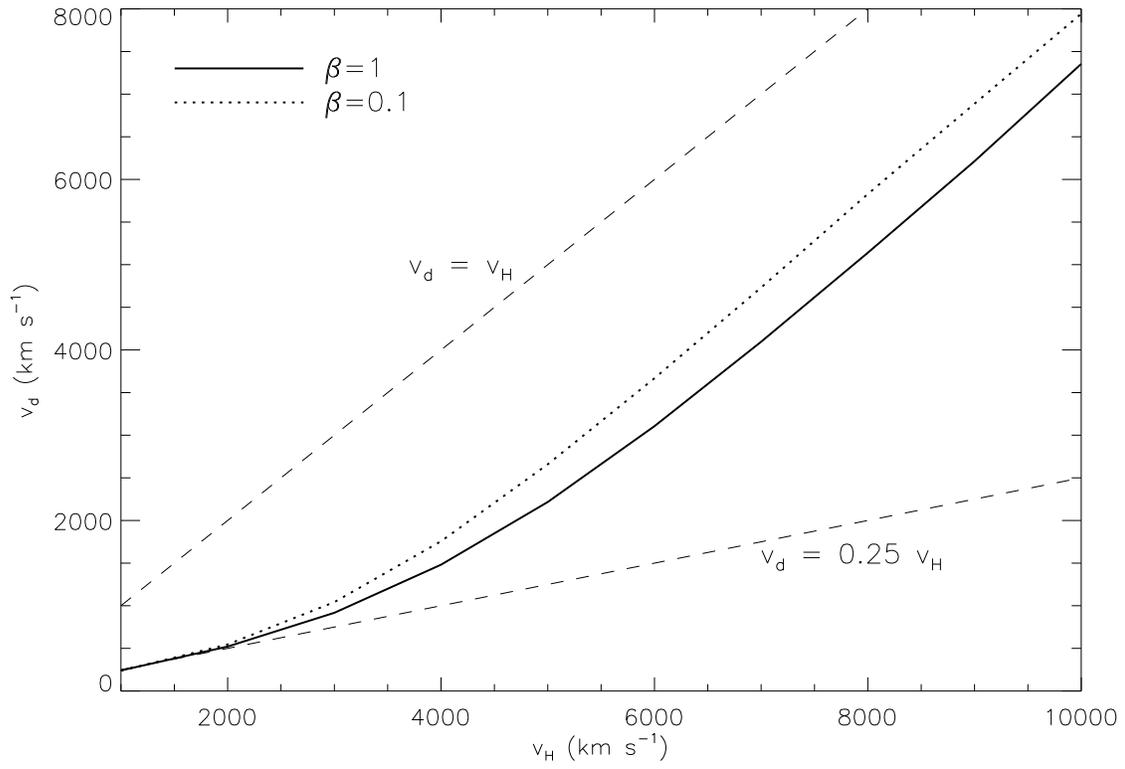}
\end{center}
\caption{Examples of the drift velocity, $v_d$, for $\beta = 0.1$ and 1.  The lower and upper bounds are shown for $1/4 \lesssim v_d/v_{\rm{H}} \lesssim 1$.}
\label{fig:drift_velocity}
\end{figure}

If we express the transfer function in the form
\begin{equation}
P(x,x_0) = P_0(x_0) ~\exp{\left[-\frac{x-x_0}{l_d(x,x_0)}\right]},
\end{equation}
then the drift length is
\begin{equation}
l_d(x,x_0) = (x - x_0) ~\left(\alpha \ln{\left[ \frac{C_1 + C_2 \exp{\left(C_3 x\right)}}{C_1 + C_2 \exp{\left(C_3 x_0\right)}} \right]} \right)^{-1}.
\end{equation}
We also note that for $f_{\rm{ion}} \gtrsim 0.9$, $l_d(x,x_0) \approx v_d/n(x_0)R_{\rm{iB}}$ is an excellent approximation for the drift length.  Examples of $v_d$ are shown in Fig. \ref{fig:drift_velocity}.


\end{document}